\pdfoutput=1
\documentclass[11pt]{article}

\usepackage[]{acl}

\usepackage{times}
\usepackage{latexsym}

\usepackage[T1]{fontenc}
\usepackage[utf8]{inputenc}

\usepackage{microtype}
\usepackage{epstopdf}
\usepackage[utf8]{inputenc}

\usepackage{hyperref}
\usepackage{multirow}
\usepackage{xstring}
\usepackage{adjustbox}

\usepackage{color}
\usepackage{longtable}
\usepackage[normalem]{ulem}
\usepackage{subfigure}
\usepackage{comment}
\usepackage{fancyvrb}
\usepackage{todonotes}

\def\MEdel#1{\bgroup\markoverwith{\textcolor{green}{\rule[0.5ex]{2pt}{1pt}}}\ULon{#1}}

\def\JNdel#1{\bgroup\markoverwith{\textcolor{red}{\rule[0.5ex]{2pt}{1pt}}}\ULon{#1}}

\def\BDdel#1{\bgroup\markoverwith{\textcolor{blue}{\rule[0.5ex]{2pt}{1pt}}}\ULon{#1}}

\title{DEPAC: a Corpus for Depression and Anxiety Detection from Speech}

\author{Mashrura Tasnim, Malikeh Ehghaghi, Brian Diep, Jekaterina Novikova \\
\{mashrura, malikeh, brian, jekaterina\}@winterlightlabs.com\\
        Winterlight Labs \\ Toronto, Canada }
\begin{document}
\maketitle
\begin{abstract}
Mental distress like depression and anxiety contribute to the largest proportion of the global burden of diseases. Automated diagnosis systems of such disorders, empowered by recent innovations in Artificial Intelligence, can pave the way to reduce the sufferings of the affected individuals. Development of such systems requires information-rich and balanced corpora. In this work, we introduce a novel mental distress analysis audio dataset DEPAC, labeled based on established thresholds on depression and anxiety standard screening tools. This large dataset comprises multiple speech tasks per individual, as well as relevant demographic information. Alongside, we present a feature set consisting of hand-curated acoustic and linguistic features, which were found effective in identifying signs of mental illnesses in human speech. Finally, we justify the quality and effectiveness of our proposed audio corpus and feature set in predicting depression severity by comparing the performance of baseline machine learning models built on this dataset with baseline models trained on other well-known depression corpora.
\end{abstract}

\section{Introduction}
Effective treatment for psychiatric diseases requires characterizing disease profiles with high accuracy. The traditional schema for diagnosis is based on clustering of non-specific physical and behavioral symptoms, which makes the diagnostic process challenging. For example, in major depressive disorder (MDD), high disease heterogeneity and lack of agreed-upon assessment standards necessitate a high degree of clinical experience and training to make an accurate diagnosis. 
Both clinician-administered and self-rated clinical assessments for MDD, such as the Hamilton Depression Scale (HAM-D) \cite{hamilton1976hamilton}, Montgomery Asberg Depression Scale (MADRS) \cite{montgomery1979new}, Beck Depression Inventory (BDI) \cite{beck1988psychometric}, and Patient Health Questionnaire (PHQ-9) \cite{lowe2004monitoring} are suboptimal in many ways. Each assess the illness through different symptom domains, have low construct validity, lack specific behavioral references, and are subjective \cite{berman1985cognitive,nemeroff2007prevalence,wakefield2013dsm}. Moreover, participants are often reluctant to fill-out the self rated assessment in regular intervals. These issues can lead to misdiagnosis, which impacts treatment timelines and can lead to poor clinical outcomes. 

In contrast, language can be an effective alternative to objectively characterize psychiatric illness. For example, emotion and cognition are both affected in MDD. As a result, depressed patients demonstrate negative emotional bias in memory, attention, and event-interpretation \cite{mathews2005cognitive}, as well as more general impairment in attention, memory, and decision-making \cite{cohen1982effort,blanco2013influence}. These effects are manifested in patients’ language in a variety of ways, for example, slowed rate of speech, volume, prosody, as well as increased use of first-person pronouns, negatively valenced speech content, and use of absolute words \cite{flint1992acoustic,fineberg2016self}. %Other studies in Schizophrenia and Bipolar disorder suggest that speech can be used to distinguish between these conditions and may be a beneficial biomarker for early relapse detection \cite{mota2017thought,mota2012speech}. 
Therefore, automated computational analysis of speech represents an excellent data source to develop digital biomarkers for mental illness. This kind of automated assessment takes only a few minutes of audio recording, therefore is less time-consuming, and would reduce burden on the individuals. However, such model development requires access to datasets of sufficient quality and size.

The recent development of speech-based computational models for measuring depression prevalence and severity has been accelerated by the introduction of Audio-Visual Emotion Recognition Challenge (AVEC) in 2013. A subset of the audio-visual depressive language corpus (AViD-Corpus) was introduced as challenge corpus for 2013 \cite{valstar2013avec} and 2014 \cite{valstar2014avec} Depression Recognition Sub-Challenge (DSC) of the event. This dataset comprises 150 recordings in German language, divided equally into training, development and test partitions. Predicting depression severity on BDI-II scale was the challenge specified task.

Another popular dataset in this area is the Distress Analysis Interview Corpus (DAIC) \cite{gratch2014distress}. It contains semi-structured clinical interviews in English language formulated to support diagnosis of psychological conditions such as anxiety, depression, and post-traumatic stress disorder. Different subsets of this dataset were used as the challenge corpus of AVEC 2016, 2017 and 2019 \cite{valstar2016avec,ringeval2017avec,ringeval2019avec} where challenge participants reported PHQ-8 scores predicted by their respective regression models.

However, the depression corpora used in previous research suffer from two vital limitations. Firstly, the small sample size in the existing depression datasets increases the risk of overfitting in the machine learning models. For example, the number of recordings in the AVEC challenges available for model training range from 50 to 189, which is far from sufficient. Secondly, the datasets in the previous works lack in linguistic variety, as they only contain a small number (only one or two) of samples per subject. To mitigate these challenges, in this work we introduce the \textbf{DEP}ression and \textbf{A}nxiety \textbf{C}rowdsourced corpus (DEPAC) as a novel dataset that is rich in the diversity of speech tasks and subjects and is tailored to capture the signs of anxiety and depression to make accurate prediction on subjects' psychological state. We also present a set of acoustic and linguistic features extracted from the corpus which incorporates domain knowledge of clinical and machine learning experts. Finally, we benchmark our dataset with several baseline machine learning models that use this set of features, to show that this novel dataset is well-suited for the machine learning-based methods with the goal of generating speech biomarkers for depression.

\section{DEPAC Corpus}
\label{Sec: data_collection}
The DEPAC corpus introduced in this work was collected with the goal of gathering a large training dataset to identify candidate speech and language features that are specific to a given psychiatric disease. Data collection for the corpus was approved by the Institutional Review Board (IRB). This is a proprietary dataset, collected via crowdsourcing and consists of a variety of self-administered speech tasks. The participants completed these tasks using Amazon Mechanical Turk\footnote{\url{https://www.mturk.com}} (mTurk), a platform where individuals are paid to complete short tasks online \cite{paolacci2010running}. The speech samples were then manually transcribed and compiled along with participant demographic information into the final corpus.
\subsection{Platform and Instrumentation}
Once recruited for this study via mTurk, participants were able to remotely complete a range of tasks including surveys and responding to audio prompts. Participants were required to have:
\begin{enumerate}
  \setlength{\itemsep}{1pt}
  \setlength{\parskip}{0pt}
  \setlength{\parsep}{0pt}
    \item A desktop or laptop computer
    \item A working microphone
    \item Chrome or Mozilla Firefox browser
\end{enumerate}
Amazon facilitated payment between the experimenter and the participant.

% \begin{figure}[h]
%     \centering
%     \includegraphics[width=1\linewidth]{figures/01_Family_in_the_kitchen_web.png}
%     \caption{`Family in the kitchen' image used for the picture description task.}
%     \label{fig:pic_desc}
% \end{figure}

\subsection{Recruitment and Screening}

Participation in the study was voluntary. Participant eligibility was configured to only permit individuals located in Canada and the United States. Amazon verified the location of participants by confirming their address and associated credit card. Locations were used to assess eligibility only.

The platform also restricted participation to individuals with an mTurk approval rating of at least 95\%. This preliminary criterion attempted to ensure that participation was restricted to those who had historically consistently followed task instructions.

During the study, participants saw a short description of the task, the approximate length of the task (5 to 8 minutes, depending on the condition they were randomly placed into), and the per-minute payment for their time. Participants were compensated at a rate of \$0.16 per minute. This is well above the average payment rate for mTurk tasks and above the recommended rate of \$0.10/minute \cite{chandler2016conducting}.

As part of our exclusion criteria, individuals with a history of chronic alcohol or drug dependency within the past 5 years, as well as participants with clinically significant vision or hearing impairment, were excluded from the study.

\subsection{Transcription and Quality Assurance}
Each audio sample gathered from the mTurk platform was assigned to a trained transcriptionist to follow the protocols and annotation formats detailed in the CHAT manual \cite{macwhinney2000childes} that was used to transcribe TalkBank, which is the largest open repository of spoken data \cite{macwhinney2007talkbank}. The transcriptionists annotated via an internally developed tool where they had access to the recording and a platform for transcribing the content of the audio file, separating the audio file into utterances, and performing quality assurance. Samples with minor audio issues not impacting the transcriptionist's ability to produce an accurate transcript were processed as normal. Samples that could not be properly transcribed due to excessive background noise, poor audio quality, or other external issues such as the presence of multiple speakers in the file were tagged as unusable and were omitted from the corpus. In total, 91 samples out of 2765 collected samples were tagged as such and omitted.

\subsection{Demographic Data Collection}
Upon consenting, participants were asked to indicate whether they are native English speakers (i.e., whether they learned the English language before the age of 5 years old). They were also asked to indicate their age, gender, and education level.

\subsection{Speech tasks}
\label{sec:speech_task}
During each recording session, the subjects completed the following standard tasks, selected to elicit speech patterns that can be analyzed for acoustic and linguistic features that correlate to psychiatric state:
\begin{figure}[h]
    \centering
    \includegraphics[width=1\linewidth]{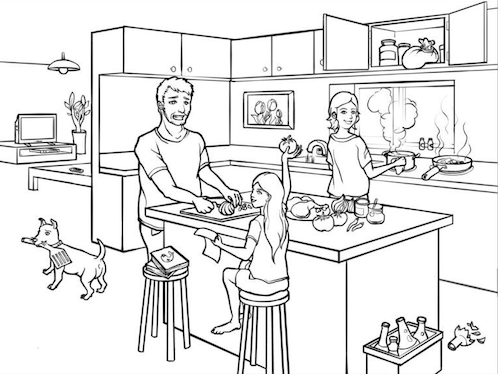}
    \caption{`Family in the kitchen' image used in the picture description task.}
    \label{fig:pic_desc}
\end{figure}

\begin{table}[t]
\vspace{0.5em}
\begin{adjustbox}{max width=1\linewidth, center}
\renewcommand{\arraystretch}{1.3}
    \centering
    \begin{tabular}{|p{0.5\linewidth}|l|l|l|}
        \hline
         \textbf{Criteria} & \textbf{AVEC \newline 2013, 2014} & \textbf{DAIC-WoZ} & \textbf{DEPAC (our)} \\
        \hline
        {Language} & German & English & English\\
        \hline
        {\# of speech tasks} & 2 & 1 & 5\\
        \hline
        {\# of samples total / per subj.} & 150 / 2 & 189 / 1 & 2674 / 5\\
        \hline
        {Depression scale} & BDI-II & PHQ-8 & PHQ-9\\
        \hline
        {Anxiety scale} & - & - & GAD-7\\
        \hline
        {Avg. depression score} & 15.34($\pm$ 12.13) & 6.65 ($\pm$ 6.11) & 6.56 ($\pm$ 5.56)\\
        \hline
        {Depression score range in the corpus} & 0-45 & 0-23 &  0-27\\
        \hline
        %\textbf{Max depression score} & 45 & 23 & 27\\
        %\hline
    \end{tabular}
\end{adjustbox}
    \caption{Description of our DEPAC dataset and its comparison to existing depression/anxiety corpora.}
    \label{tab:depac}
\end{table}

\begin{figure*}%
\centering
\subfigure[Age distribution]{%
\label{fig:age-dist}%
\includegraphics[width=0.7\linewidth]{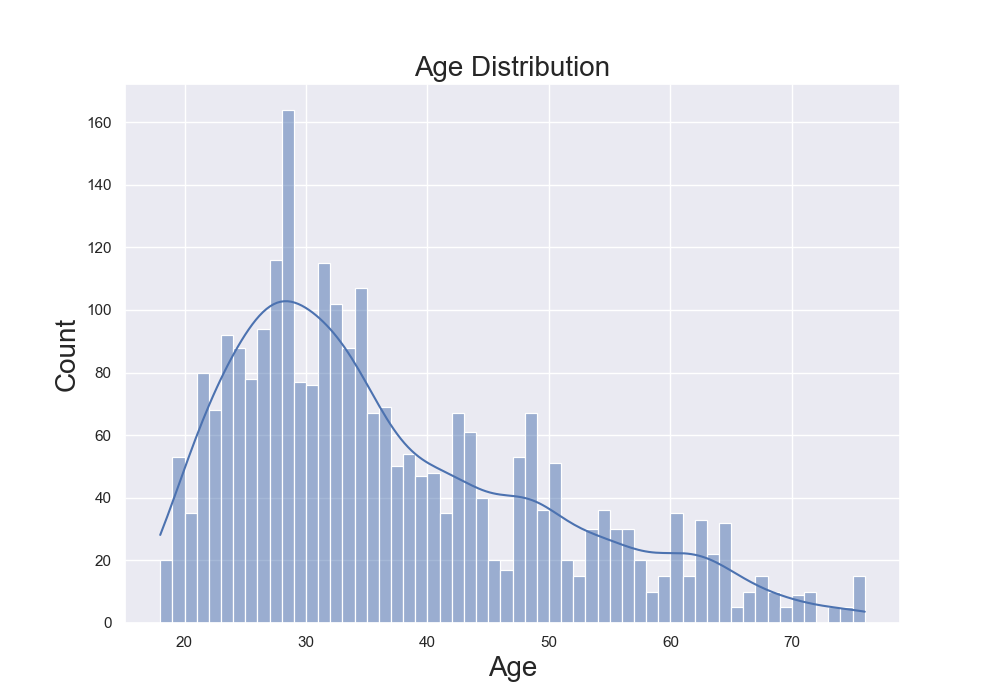}}
\quad
\subfigure[Distribution of the formal years of education]{%
\label{fig:education-dist}%
\includegraphics[width=0.7\linewidth]{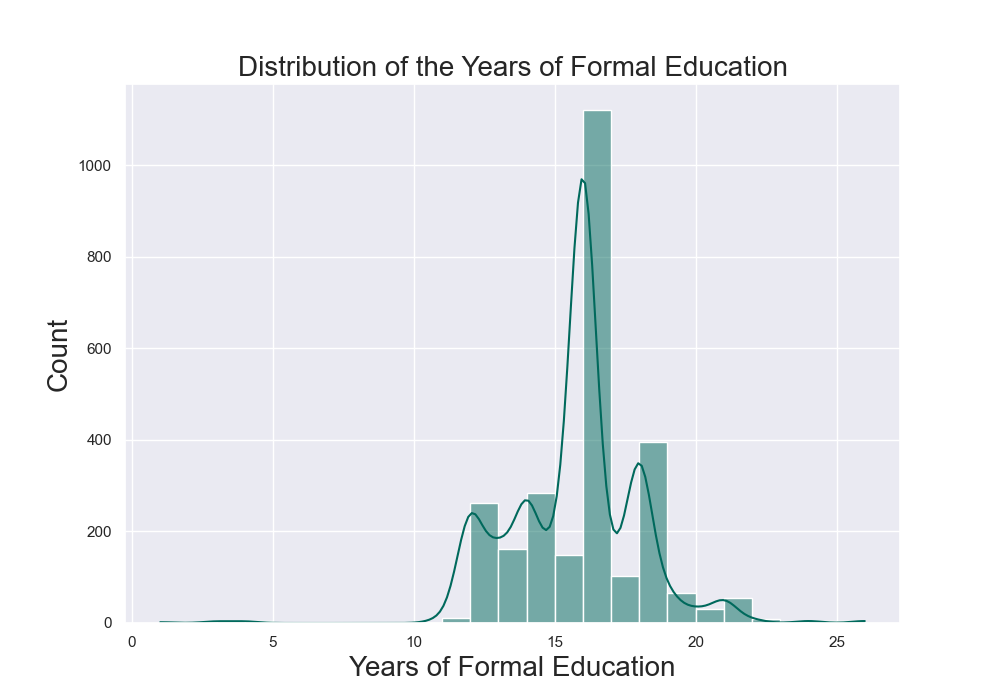}}%
\quad
\subfigure[Distribution of PHQ-8 scores in AVEC 2019 and DEPAC dataset by gender]{%
\label{fig:PHQ_dist}%
\includegraphics[width=0.8\linewidth]{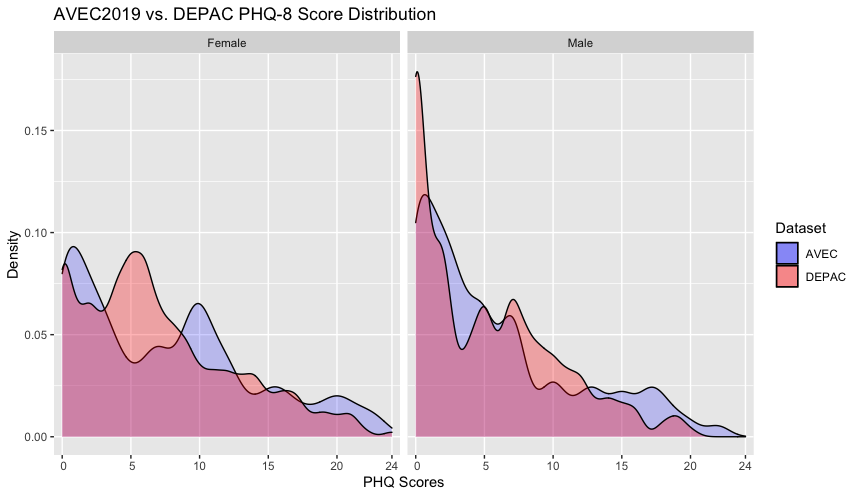}}%
\caption{Distribution of the participants' demographics in mTurk Study.}
\label{fig:demographic-info}
\end{figure*}

\begin{itemize}
  \setlength{\itemsep}{1pt}
  \setlength{\parskip}{0pt}
  \setlength{\parsep}{0pt}
  \item \textbf{Phoneme Task:} Participants were asked to sustain a phoneme sound (e.g., /ā/) for as long as they could, up to one minute. They could cease making the sound whenever they choose. Due to difficulty in finding voiced parts in continuous speech, sustaining vowels would be optimal for measuring source and respiration  features (e.g., shimmer) \cite{low2020automated}.
  \item \textbf{Phonemic fluency: }Phonemic verbal fluency was evaluated using the FAS (`F', `A', `S') \cite{borkowski1967word} task (letter ``F"). This assessment has been used widely in a variety of populations, including individuals with Alzheimer's Disease (AD). The average duration of this speech task was 22.13 seconds in DEPAC dataset.
  \item \textbf{Picture description}: A static image depicting an event was presented to the subject, and they were asked to describe what is happening in their own words. The average length of picture-based narratives was 46.60 seconds. Tasks of this type have been shown to be good proxies for spontaneous discourse \cite{giles1996performance}. Picture description was found to be an effective speech task in evoking situations that required more cognitive effort and caused noticeable changes in speech for detecting depression \cite{jiang2017investigation}. In this study, a proprietary image `Family in the kitchen' (Figure \ref{fig:pic_desc}) was used, which was designed to match the `Cookie theft' picture \cite{goodglass2001bdae} in style and content units. The picture was a line drawing of an everyday scene, containing three to four characters, two salient action items (e.g., broken bottle, or steaming pot), and a similar number of object units (20-25), action items (9-10) and locations (2) \cite{forbes2005detecting}. Our core design guidelines to develop this picture are listed in \ref{apd:pic_guidelines}.

  \item \textbf{Semantic fluency: } Participants were asked to list as many positive future experiences as they can within one minute. They were given time parameters to guide them, such as future events predicted to happen within three weeks, within one month, within one year, and so on. They were allowed to describe as little or as much as they choose. Performance on verbal fluency tasks are found to correlate with executive deficits caused by depression \cite{fossati2003qualitative}. The length of speech in this task was 43.76 seconds on average.
  \item \textbf{Prompted narrative: } Participants were asked to describe an event, interest, or hobby based on a single prompt, e.g., “Describe your day”, “Describe a travel experience” and “Describe a hobby you have”. Participants were allowed to describe as much or as little as they choose. Narrative speech provides an opportunity to elicit speech containing the linguistic structures and acoustic information that is known to contain indicators of depression \cite{trifu2017linguistic}. The average duration of the prompted speech in the collected dataset was 45.34 seconds.
\end{itemize}

\subsection{Clinical Assessments}
The following two mental health assessment questionnaires were completed by the participants after the recording session:

\textbf{Patient Health Questionnaire (PHQ-9):} The PHQ-9 is a well established 3-point self-rated measure for\textbf{ depressive symptoms} that has been validated against clinician rated measures \cite{kroenke2001phq}. It contains 9 questions which correspond to the core criteria of the Diagnostic And Statistical Manual of Mental Disorders (DSM) for depression. Scores on this scale range from 0 to 27 with diagnostic cut-off thresholds for depression severity. Scores less than 5 represent the individuals with no depression; individuals with a mild or subthreshold depressive disorder are reflected by scores from 5 to 9; scores between 10 and 14 indicate moderate severity level of depression, and scores 15 or higher signify major depressive disorder in the participants \cite{kroenke2001phq}.

\textbf{Generalized Anxiety Disorder - 7 (GAD-7):} The GAD-7 is a popular self-rated measure of general \textbf{anxiety symptoms} that is scored from 0 to 21 \cite{spitzer2006brief}. It has been validated against clinical diagnosis and has been shown to be robust as a screening tool and a continuous measure of symptom severity. Scores of 10 or above indicate a reasonable threshold for detecting individuals with generalized anxiety disorder. Similar to the levels of depressive disorder in PHQ-9,  5, 10, and 15 are the cut points on the GAD-7 scale to classify anxiety severity level into minimal, mild, moderate and severe groups \cite{spitzer2006brief}.

\subsection{Corpus Composition}
The dataset consists of 2,674 audio samples collected from 571 subjects (Table~\ref{tab:depac}). 54.67\% of the study subjects are female and 45.33\% are male. The age of the subjects ranges between 18 and 76, and they received 1 to 26 years of formal education.

Figure \ref{fig:demographic-info} illustrates the demographic distribution of the mTurk study. %Figures \ref{fig:age-dist} and \ref{fig:education-dist} represent the distribution of the subjects' age and years of formal education in this dataset. 
The age distribution is shifted toward the left around its average value, which is equal with 36.85, indicating that most of the dataset is made up of young or middle-aged adults (Figure~\ref{fig:age-dist}). Moreover, it is witnessed in the education level distribution plot that the most of the participants received higher education, with on average around 15 years of formal education (Figure~\ref{fig:education-dist}). 

Figure \ref{fig:GAD_PHQ_dist} (Appendix \ref{sec:score_dist}) demonstrates that the distribution of both GAD-7 and PHQ-9 scores are skewed-right, representing that the majority of the dataset is composed of either no or subthreshold level of the disorders. In addition, the number of samples with moderate to severe level of both disorders are higher among women compared with men.

\section{Feature Sets}
In this section, we introduce a set of hand-crafted features extracted from the DEPAC audio records and the associated transcripts. The set of features comprises various linguistic and acoustic features that have been found in previous psychiatric literature to be effective in detection of depression and anxiety from spoken language \cite{low2020automated, smirnova2018language}.

\subsection{Acoustic Features: }
We extracted 220 acoustic features from each audio sample. The feature set includes:

\onecolumn
\begin{center}
\begin{longtable}{|p{0.2\linewidth}|p{0.7\linewidth}|}
    \multicolumn{2}{c}{\textbf{Generic Linguistic Features}}\\
    \hline
    \textbf{Feature Category} &  \textbf{Description}\\
    \hline
    \endfirsthead
    % \multicolumn{2}{c}%
    % {\tablename\ \thetable\ - \textit{List of linguistic features in our conventional feature set -- Continued from previous page}} \\
    % \hline
    % \textbf{Feature Category}  &  \textbf{Description}\\
    % \hline
    % \endhead
    % \hline \multicolumn{2}{r}{\textit{Continued on next page}} \\
    % \endfoot
    % \hline
    % \endlastfoot
    Discourse mapping (18) & \textbf{Utterance distances} and \textbf{speech-graph} \cite{mota2012speech} features extracted from the graph representation of the transcripts.\\
    \hline
    Local coherence (15) & Average, maximum, and minimum similarity between Word2Vec \cite{mikolov2013distributed} representations of the successive utterances.\\
    \hline
    Lexical complexity and richness (103) &
    \textbf{Vocabulary richness}: Such as Brunet's index \cite{brunet1978vocabulaire} and Honore's statistic \cite{honore1979some}.
    
    \textbf{Psycholinguistics norms}: 
    Average norms across all words, nouns only and verbs only for imageability, age of acquisition, familiarity \cite{stadthagen2006bristol} and frequency (commonness) \cite{brysbaert2009moving}.
    
    \textbf{Grammatical constituents}: The constituents comprising the parse tree in a set of Context-Free Grammar (CFG) features.\\
    \hline
    
    Syntactic complexity (143) & \textbf{Constituency-parsing based features}: Scores based on the parse tree \cite{chae2009predicting} (e.g., the height of the tree, the statistical functions of Yngve depth (a measure of embeddedness) \cite{yngve1960model}, and the frequencies of various production rules\cite{chae2009predicting}).
    
    \textbf{Lu's syntactic complexity features}: Metrics of syntactic complexity suggested by \cite{lu2010automatic} such as the length of sentences, T-units, and clauses, etc.
    
    \textbf{Utterance length}: Average, maximum and minimum utterance length.\\
    
    \hline
    Utterance cohesion (1) & Number of switches in verb tense across utterances divided by total number of utterances.\\
    \hline
    
    Sentiment (9) & Variables such as valence, arousal, and dominance scores \cite{warriner2013norms} for all words and word types describing the sentiment of the words used.\\
    \hline
    Word finding difficulty (11) & \textbf{Pauses and fillers}:
     Variables like speech rate, hesitation, duration of words and number of filled (e.g., um, uh) and unfilled pauses as signs of word finding difficulty, which result in less fluid or fluent speech \cite{pope1970anxiety}.
    
    \textbf{Invalid words}: Not in Dictionary (NID) indicating proportion of words not in the English dictionary.\\
    \hline
    \multicolumn{2}{c}{\textbf{Task-Specific Linguistic Features}}\\
    \hline
    \textbf{Speech Task}  &  \textbf{Description}\\
    \hline
    Phonemic Fluency (2) & Includes the raw number of words starting with the correct letter with/without explicit filtering out of proper nouns by their Part of Speech (POS) tags.\\
    \hline
    Picture Description (25) & \textbf{Global coherence}: Average, minimum and maximum cosine distance between GloVe \cite{pennington2014glove} word vector representation of each utterance and its closest content unit centroid utterances.\newline
    
    \textbf{Information units}: The number of objects, subjects, locations and actions used to measure the number of items correctly named in the picture description task.\\
    \hline
    Semantic Fluency (1) & Includes the raw number of words of the correct category.\\
    \hline
    \caption{List of linguistic features in our conventional feature set. The number of features in each subtype is shown in the parentheses.}
    \label{tab:linguistic-features}
\end{longtable}
\end{center}

\clearpage
\twocolumn

\begin{itemize}
  \setlength{\itemsep}{1pt}
  \setlength{\parskip}{0pt}
  \setlength{\parsep}{0pt}
    \item \textbf{Spectral features: } Intensity (auditory model based), MFCC 0-12, Zero-Crossing Rate (ZCR)
    \item \textbf{Voicing-related features: }Fundamental frequency $(F_0)$, Harmonic-to-Noise Ratio (HNR), shimmer and jitter, durational features, pauses and fillers, phonation rate
\end{itemize}
Statistical functionals including minimum, maximum, average, and variance were computed on the low-level descriptors. Additionally, skewness and kurtosis were calculated on MFCCs, first and second order derivatives of MFCCs, and Zero Crossing Rate (ZCR) \cite{low2020automated} (Table \ref{tab:long-acoustic-features} in appendix elaborates on detailed descriptions of these features as well as previous literature motivating their selection as the indicators of psychiatric conditions).

A Python implementation of Praat phonetic analysis toolkit \cite{boersma2001speak} has been used to extract the majority of these features. The MFCC features and their functionals were computed using \verb|python_speech_features|\footnote{\url{https://pypi.org/project/python\_speech\_features/}} library.

\subsection{Linguistic Features:} 
We also applied standard natural language processing libraries (e.g., spaCy\footnote{\url{https://spacy.io/}} and Stanford Parser\footnote{\url{https://nlp.stanford.edu/software/lex-parser.html}}) to extract 300 generic and 28 task-specific linguistic features from the associated transcripts of the audio files (Table \ref{tab:linguistic-features}). For simplification, we classified the generic features into the categories including discourse mapping, local coherence, lexical complexity and richness, syntactic complexity, utterance cohesion, sentiment, and word finding difficulty (the selection motivations of our linguistic features are explained in Appendix \ref{apd:feat-desc-long}, Table \ref{tab:long-linguistic-features}).

\section{Intended Usage}
The study aimed to collect a high quality training dataset with the intention of developing a speech-based digital biomarker for the psychiatric diseases of depression and anxiety. The dataset is well-suited for exploratory analysis involving statistical and machine learning methods to generate potential speech biomarkers and test their validity. In Section \ref{Sec:Baseline}, we present the baseline models to predict depression severity using this dataset, that can be used as benchmarks for the future research.

\section{Baseline Models for Depression Analysis}
\label{Sec:Baseline}
\subsection{Data Preprocessing}
\paragraph{Standardization: }
Once the acoustic and linguistic features were extracted from the data records, we standardized them using z-scores, i.e., subtracting the mean and dividing by standard deviation. The standard score of a sample $x$ of feature $f_i$ is calculated as:

\begin{equation}
y = \frac{x-\mu}{\sigma}
\end{equation}
here $\mu$ and $\sigma$ are the mean and standard deviation of the values of $f_i$ in all training samples.

% My counts table
\begin{table}[t]
\vspace{0.3em}
\begin{adjustbox}{max width=1\linewidth, center}
\renewcommand{\arraystretch}{1.3}
    % \begin{adjustbox}{max width=0.8\textwidth, center}
    \renewcommand{\arraystretch}{1.3}
    \centering
    \begin{tabular}{|p{0.2\linewidth}|r|r|r|}
        \hline
        \textbf{Range of scores} & \textbf{AVEC PHQ-9} & \textbf{DEPAC PHQ-9} & \textbf{DEPAC GAD-7} \\ 
        \hline
        [0 - 5) & 77 & 240 & 261 \\
        \hline
        [5 - 10) & 36 & 178 & 152 \\
        \hline
        [10 - 15) & 26 & 84 & 87 \\
        \hline
        [15 - 20) & 17 & 40 & 45 \\ 
        \hline
        [20 - 27] & 7 & 10 & 7 \\
        \hline

    \end{tabular}
    \end{adjustbox}
    \caption{Counts for the PHQ-8/GAD-7 scores in AVEC and DEPAC datasets}
    \label{tab:severity_counts}
    % \end{adjustbox}
\end{table}

\subsection{Model Training}
To compare the efficacy of different modalities in predicting depression, we trained a combination of linear and non-linear Machine Learning (ML) models: Support Vector Regressors (SVR), Linear Regression (LR), and Random Forest Regressor (RF) separately on the following feature categories:
\begin{enumerate}
  \setlength{\itemsep}{0pt}
  \setlength{\parskip}{0pt}
  \setlength{\parsep}{0pt}
    \item Demographic features (i.e., age, gender, and education)
    \item Acoustic features
    \item Linguistic features
\end{enumerate}

We further investigated the effectiveness of each speech task for predicting depression severity on the PHQ-8 scale. The main reason for excluding the last question in PHQ-9 questionnaire was to make the results comparable to the performances with AVEC 2016 \cite{valstar2016avec} and AVEC 2019 \cite{ringeval2019avec} baselines, which are reported on PHQ-8 scale. The audio samples in AVEC challenges are subsets of Distress Analysis Interview Corpus (DAIC-WoZ) \cite{gratch2014distress}, which includes interviews of the participants conducted by a virtual agent. The length of the speech samples of the DAIC-WoZ dataset range from 5 to 25 minutes, including both participants' and interviewer's speech. 

Figure \ref{fig:PHQ_dist} compares how the PHQ-8 scores are distributed in male versus female participants in AVEC 2019 and DEPAC datasets. Higher PHQ scores indicates the higher depression severity in the subjects. The distributions are skewed-right both for the male and female participants, representing that the majority of both datasets is composed of either no or mild level of depression. The number of samples in each level of depression in each of the two datasets is summarized in Table \ref{tab:severity_counts}.

To validate the comparison of our models' performance with the ones trained on the AVEC datasets, we performed independent t-test on the PHQ-8 score distribution of the DEPAC dataset and AVEC 2019 corpus. The outcome of the test showed that the two datasets do not exhibit significant differences $(t=0.65, p > 0.05)$%. The test score indicated that, 
and as such, these two datasets are similar enough to compare the performance of the baseline ML models.

Compared with previous datasets, our dataset is enriched with a greater variety of speech tasks. Thus, in addition to an analysis using data from all the included tasks, we evaluate models trained on task-subsets of the corpus and report their performance in predicting depressive disorder. Each model is evaluated with regard to the Mean Absolute Error (MAE) and Root Mean Square Error (RMSE) scales, following the baseline set by AVEC challenge \cite{valstar2016avec}, \cite{ringeval2017avec}. The performance metrics are described in Appendix \ref{apd:metrics}.

We trained an SVR model on the combination of acoustic and linguistic features extracted from all five speech tasks (See Section \ref{sec:speech_task}), and also separately on each of the speech (See Table \ref{tab:task}).

For all the experiments, all model hyperparameters were set to their default values as on the Scikit-learn implementation \cite{scikit-learn}. Models were trained using grouped 10-fold cross validation, where samples from the same individual do not appear in both the training folds and test fold. All results are reported as the mean MAE/RMSE scores across the 10 folds.

\subsection{Baseline Model Result and Discussion}

We present and discuss the results of baseline model training across different modalities of input features, i.e. demographic, acoustic and linguistic, as well as across five different speech tasks, using DEPAC speech data.

\noindent \textbf{Model Performance across Modalities:} Among the three modalities, SVR model trained on demographic features performs the best, achieving the lowest MAE and RMSE, followed by the SVR model trained on linguistic features. Both acoustic and linguistic baseline models attain less than 20\% MAE in the range of scores (0 to 24). Marginal deviation of both MAE and RMSE between acoustic and linguistic models suggests that these two modalities are effective for the task of recognizing signs of depression from speech. It is noteworthy that, the audio files did not undergo any pre-processing or enhancement before extracting the acoustic features. Yet, models trained on acoustic features exhibit competitive performance with the linguistic model, indicating that the quality of the recordings is sufficient and is a valuable foundation for future research.

\begin{table}[t]
\vspace{0.3em}
\begin{adjustbox}{max width=0.9\linewidth, center}
% \renewcommand{\arraystretch}{1.5}
    % \begin{adjustbox}{max width=0.8\textwidth, center}
    \renewcommand{\arraystretch}{1.3}
    \centering
    \begin{tabular}{|p{0.25\linewidth}|p{0.3\linewidth}|r|r|}
        \hline
        \textbf{Features} & \textbf{Algorithm} & \textbf{RMSE} & \textbf{MAE} \\ 
        \hline  
        \multirow{3}{*}{Demographic} & LR  & 6.94	& 5.18 \\
        \cline{2-4}
        & RF  & 6.34 & 4.93\\
        \cline{2-4}
        & SVR & \textbf{5.20} & \textbf{4.06}\\
        \hline
        \multirow{5}{*}{Acoustic}  & LR  & 7.51 & 5.86 \\
        \cline{2-4}
        & RF & 5.41 & 4.41 \\
        \cline{2-4}
        & SVR & 5.48 & 4.40 \\
        \cline{2-4}
        & AVEC 2016 baseline \cite{valstar2016avec} & 7.78 & 5.72\\
        \cline{2-4}
        & AVEC 2019 baseline \cite{ringeval2019avec} & 8.19 & -\\
        \hline 
        \multirow{3}{*}{Linguistic} & LR & 5.72 & 4.60 \\
        \cline{2-4}
        & RF & 5.40 & 4.37\\
        \cline{2-4}
        & SVR & 5.37 & 4.24 \\
        \hline

    \end{tabular}
    \end{adjustbox}
    \caption{Regression results of the models predicting PHQ-8 score on different categories of features. Bold indicates the best performance.}
    \label{tab:performance}
    % \end{adjustbox}
\end{table}

In terms of predicting PHQ-8 scores, our baseline models perform substantially better than the baseline models specified by challenge organizers of AVEC 2016 \cite{valstar2016avec} and AVEC 2019 \cite{ringeval2019avec} (Table \ref{tab:performance}), despite the shorter length of samples than the AVEC corpus, which justify the robustness of the hand-curated acoustic features introduced in this work, as well as the quality of the dataset. 

\begin{table}[t]
\vspace{0.3em}
\begin{adjustbox}{max width=0.9\linewidth, center}
% \renewcommand{\arraystretch}{1.5}
    % \begin{adjustbox}{max width=0.8\textwidth, center}
    \renewcommand{\arraystretch}{1.3}
    \centering
    \begin{tabular}{|p{0.4\linewidth}|r|r|}
        \hline
        \textbf{Speech task} & \textbf{RMSE} & \textbf{MAE} \\ 
        \hline  
        Phoneme Task  & 5.49 & 4.32 \\
        \hline
        Phonemic fluency & 5.44 & 4.31 \\
        \hline
        Picture description & 5.36  & 4.25 \\
        \hline
        Positive fluency & \textbf{5.19} & \textbf{4.11}\\
        \hline
        Prompted narrative & 5.30 & 4.20 \\
        \hline
        All tasks & 5.38 & 4.27 \\
        \hline

    \end{tabular}
    \end{adjustbox}
    \caption{Regression results of SVR models predicting PHQ-8 score on different speech tasks. Bold indicates the best performance.}
    \label{tab:task}
    % \end{adjustbox}
\end{table}

Surprisingly, the SVR model using only demographic features outperforms both acoustic and linguistic models (Table \ref{tab:performance}). This demographic information was previously found to be highly correlated to one's level of depression in literature \cite{akhtar2007relation}. However, in real-world application, the demographic model may not be completely reliable due to ambiguity of these features.  

\noindent \textbf{Model Performance across Speech Tasks:} In our task-specific analysis, comparatively lower RMSE and MAE are scored by models trained on picture description, positive fluency and prompted narrative than the phoneme task, phonemic fluency and all tasks combined. The possible reason behind this observation is that the picture description, positive fluency and prompted narrative tasks produce longer audio samples, resulting in more informative acoustic and linguistic features, leading to more accurate models. This observation shows that long recordings of narrative tasks can be rich sources of markers to predict depressive disorder from speech.

\section{Conclusion}
In this work, we introduce DEPAC, a rich audio dataset for mental health research which is labelled with scores on standard scales of two highly prevalent mental disorders: PHQ-9 scores for depression and GAD-7 scores for anxiety assessment. The dataset offers a remarkably larger sample size in comparison to other publicly available corpora. One other source of novelty of the presented corpus is its richness in the diversity of speech tasks and participants with various degrees of education, genders, and age groups. We also introduce a hand-curated set of acoustic and linguistic features incorporating domain knowledge of clinical and ML experts, which are used as the predictors of models for quantifying depression severity. We present the performance of baseline models in prediction of depression severity level, that can be applied by future researchers as a benchmark. Our baseline models achieve competitive performance when compared to the AVEC 2016 and AVEC 2019 baseline models and demonstrate the quality of the DEPAC dataset and effectiveness of our proposed feature set in measuring depression severity.

\bibliography{anthology,custom}
\bibliographystyle{acl_natbib}

\clearpage

\appendix
\newpage

\section{Appendix}
\label{sec:appendix}

\subsection{Picture Design Guidelines}
\label{apd:pic_guidelines}
To develop the 'Family in the kitchen' image (Figure \ref{fig:pic_desc}) for our picture description task, we used the core design principles \cite{patel2014park} described below: 
\begin{enumerate}
  \item Image content breakdown should contain:
    \begin{enumerate}
        \item \textbf{2 scenes/locations} (e.g., kitchen, or living room)
        \item \textbf{20 to 25 objects} (e.g., knife, pan, or cupboard)
        \item \textbf{9 to 10 actions} (e.g., chop, cook, steam, or fall)
        \item \textbf{3 to 4 people/subjects} (e.g., dad, dog, mom, or daughter)
        \item \textbf{2 “dangerous” elements} (e.g., broken bottle, or steaming pot)
    \end{enumerate}
  \item Images should display \textbf{relationships between components} in a scene.
  \item Images should depict \textbf{familiar themes}, but they must be accessible to adults with diverse cultural backgrounds, sexual orientations, and various socioeconomic strata.
 \item Images should be designed appropriately for \textbf{older adults} with varied levels of visual impairment.
 \item Images should provoke spontaneous discourse useful in \textbf{diagnosis and assessment} of mental health conditions. It should:
    \begin{enumerate}
        \item Elicit tokens whose labels \textbf{span the phonetic range} useful in diagnosing motor speech difficulty.
        \item Elicit tokens whose labels \textbf{span lexical norms} (varying age of acquisition (AoA), familiarity, and imageability). Representing a varied range of lexical norms allows for using the same image to test speakers with varying degrees of cognitive and language impairment.
        \item \textbf{Contain sub-scenarios} \cite{patel2014park} which would be useful generally for generating longer speech samples, and specifically in assessing discourse structure (e.g., coherence, repetition, trajectory (what order are the sub-scenarios described in), content units (which sub-scenarios are mentioned and which left out), reasoning/inferences (e.g., interconnections and causation between the sub-scenarios)).
    \end{enumerate}
\end{enumerate}

The goal of these guidelines was to keep the content generalizable across diverse cultures and to control the similarity with the `Cookie theft' \cite{goodglass2001bdae} image in lexico-syntactic complexity and the amount of information content units.

\subsection{Distribution of Assessment Scores}
\label{sec:score_dist}
\begin{figure}[h]
    \centering
    \subfigure[Distribution of PHQ-9 scores per gender]{{\includegraphics[width=0.45\textwidth]{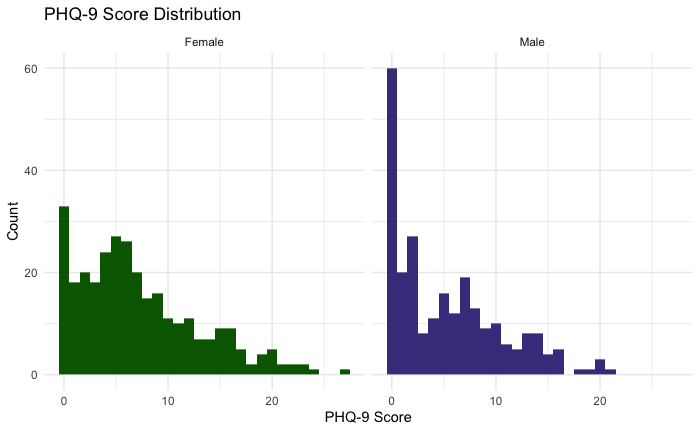}\label{fig:PHQ9} }}
    
    \subfigure[Distribution of GAD-7 scores per gender]{{\includegraphics[width=0.45\textwidth]{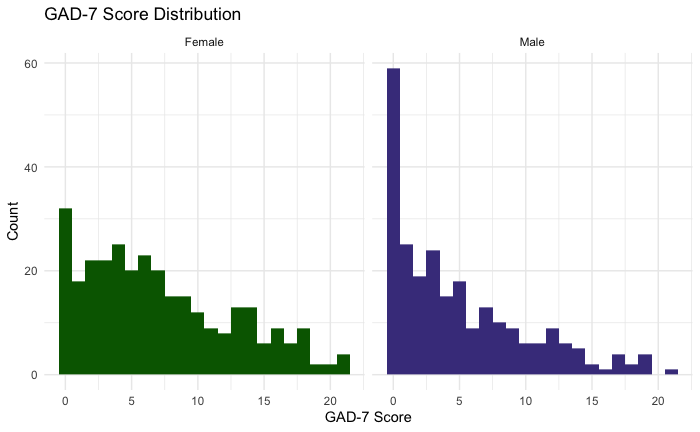}\label{fig:gad7} }}
    \caption{Distribution of the participants' PHQ-9 and GAD-7 scores in mTurk Study.}
    \label{fig:GAD_PHQ_dist}
\end{figure}

\subsection{Feature Selection Motivations}
\label{apd:feat-desc-long}
The prior studies supporting the choice of our conventional feature set are described in Table \ref{tab:long-linguistic-features} and \ref{tab:long-acoustic-features}. Table \ref{tab:long-acoustic-features} displays the selection motivations of our acoustic features derived from the audio files, including spectral and energy related as well as voicing related features. In addition, Table \ref{tab:long-linguistic-features} represents the motivations behind the choice of the generic and task-specific linguistic features extracted from the associated transcripts. 

\onecolumn
\begin{center}
\begin{longtable}{|p{0.2\linewidth}|p{0.7\linewidth}|}
    % \caption{Support literature motivating the selection of the linguistic features in our conventional feature set.}\\
    \multicolumn{2}{c}{\textbf{Generic Linguistic Features}}\\
    \hline
    \textbf{Feature Category} & \textbf{Motivations}\\
    \hline
    \endfirsthead
    % \multicolumn{2}{c}%
    % {\tablename\ \thetable\ - \textit{Support literature motivating the selection of the linguistic features in our conventional feature set.}} \\
    % \hline
    % \textbf{Feature Category} & \textbf{Motivations}\\
    % \hline
    % \endhead
    % \hline \multicolumn{2}{r}{\textit{Continued on next page}} \\
    % \endfoot
    % \hline
    % \endlastfoot
    Discourse mapping & Techniques to formally quantify utterance similarity and disordered speech via distance metrics or graph-based representations have been used to differentiate speech from those suffering from various other mental health issues that are known to affect speech production \cite{mota2012speech, fraser2016linguistic}.\\
    \hline
    Local coherence & Coherence and cohesion in speech is associated with the ability to sustain attention and executive functions \cite{barker2017cohesive}. Depression and anxiety are both known to impair such cognitive processes \cite{leung2009selective, snyder2014opposite}.\\
    \hline
    Lexical complexity and richness & Language pattern changes in particular related to the irregular usage patterns of words of certain grammatical categories such as pronouns or verb tenses have been found to differentiate depression from normal fluctuations in mood from healthy individuals \cite{smirnova2018language}.\\
    \hline
    
    Syntactic complexity & Previous literature suggests that syntactic complexity of utterances, can be used to predict symptoms of depression \cite{smirnova2018language}, including utterances elicited in self-administered contexts \cite{zinken2010analysis}.\\
    
    \hline
    Utterance cohesion &  Rates of verb tense use (in particular the past-tense) is known to be changed in individuals with depression. \cite{smirnova2018language}.\\
    \hline
    
    Sentiment & Emotional state and speech are connected, and sentiment scores in speech have been used to predict depression and anxiety levels in past research \cite{howes2014linguistic, zucco2017sentiment}.\\
    \hline
    Word finding difficulty & Previous work has found relationships between speech disturbance, filled, and unfilled speech of individuals with anxiety and depression \cite{pope1970anxiety}.\\
    \hline
    \multicolumn{2}{c}{\textbf{Task-Specific Linguistic Features}}\\
    \hline
    \textbf{Speech Task} & \textbf{Motivations}\\
    \hline
    Phonemic Fluency & Measures of individual performance at the phonemic fluency task \cite{borkowski1967word}.\\
    \hline
    Picture Description & Measures of individual performance at picture description task as defined in \cite{giles1996performance, jiang2017investigation}.\\
    \hline
    Semantic Fluency & Measures of individual performance at the semantic fluency task \cite{fossati2003qualitative}.\\
    \hline
    \caption{Support literature motivating the selection of the linguistic features in our conventional feature set.}
    \label{tab:long-linguistic-features}
\end{longtable}
\end{center}

\begin{table*}[htbp]
% \begin{adjustbox}{max width=0.9\textwidth, center}
\renewcommand{\arraystretch}{1.3}
    \centering
    \setlength\tabcolsep{2pt}
    \begin{tabular}{|p{0.25\linewidth}|p{0.7\linewidth}|}
    \multicolumn{2}{c}{\textbf{Spectral and Energy Related Features}}\\
    \hline
    \textbf{Feature} &  \textbf{Motivations}\\
    \hline
    Intensity (auditory model based) & Perceived loudness in $dB$ relative to normative human auditory threshold. In 1921, Emil Kraepelin recognized lower sound intensity in the voices of depressed patients \cite{kraepelin1921manic}.\\
    \hline
    MFCC 0-12 &  MFCC 0-12 and energy, their first and second order derivatives are calculated on every 16 ms window and step size of 8 ms, and then, averaged over the entire sample. MFCCs and their derivatives were included as baseline features in AVEC since 2013 \cite{valstar2013avec}, \cite{valstar2016avec}, \cite{ringeval2019avec} and found to be effective in predicting depression severity in the literature \cite{ray2019multi}, \cite{rejaibi2022mfcc}.\\
    \hline
    Zero-crossing rate (ZCR) & Zero crossing rate across all the voiced frames showing how intensely the voice was uttered. It was used as a speech biomarker of depression in previous studies \cite{bachu2008separation, shin2021detection}.\\
    \hline
    \multicolumn{2}{c}{\textbf{Voicing Related Features}}\\
    \hline
    $F_0$ & Fundamental frequency in Hz. A drop in $F_0$ and $F_0$ range indicates monotonous speech, which is common in depression \cite{low2020automated}. In addition, many studies have discovered a considerable rise in mean  $F_0$ in people suffering from social anxiety disorder \cite{gilboa2014being, galili2013acoustic}.\\
    \hline
    Harmonics-to-noise-ratio (HNR) & Degree of acoustic periodicity in dB using both auto-correlation and cross-correlation method. Decreasing HNR ratio has been found to correlate with increasing severity of depression \cite{quatieri2012vocal}.\\
    \hline
    Jitter and shimmer & Jitter is the period perturbation quotient and shimmer is the amplitude perturbation quotient representing the variations in the fundamental frequency. In previous studies, anxious patients indicated substantially higher shimmer and jitter. In addition, rise in jitter and shimmer variability was observed in subjects with major depressive disorder \cite{low2020automated}.\\
    \hline
    Durational features & Total audio and speech duration in the sample. In prior studies, depression severity increased the total duration of speech because of longer pauses resulting in lower speech to pause ratio \cite{alpert2001reflections, mundt2007voice}.\\
    \hline
    Pauses and fillers & Number and duration of short ($< 1s$), medium ($1-2 s$)  and long ($>2 s$) pauses, mean pause duration, and pause-to-speech ratio. Depression and anxiety are known to affect the rate of pauses/speech in individuals \cite{pope1970anxiety}.\\
    \hline
    Phonation rate & Number of voiced time windows over the total number of time windows in a sample.\\
    \hline
    \end{tabular}
    \caption{Support literature motivating the selection of the acoustic features in our conventional feature set.}
    \label{tab:long-acoustic-features}
% \end{adjustbox}
\end{table*}

\clearpage

\twocolumn
\subsection{Performance Metrics}
\label{apd:metrics}
Root Mean Square Error (RMSE) and Mean Absolute Error (MAE) are calculated using the formulas shown below.

\begin{equation}
    RMSE = \sqrt{\frac{\sum_{i=1}^{N} (x_i - y_i)^2}{N}}
\end{equation}

\begin{equation}
    MAE = \frac{\sum_{i=1}^{N} |x_i - y_i|}{N}
\end{equation}

In the above, $x_i$ and $y_i$ are the true and predicted scores respectively.

\end{document}